 \documentclass[10pt,preprint]{aastex}  
 \usepackage{natbib}
\def\ang{\AA}
\def\arcsec{\hbox{$^{\prime\prime}$}}

\def\gapprox{\lower.4ex\hbox{$\;\buildrel >\over{\scriptstyle\sim}\;$}}
\def\lapprox{\lower.4ex\hbox{$\;\buildrel <\over{\scriptstyle\sim}\;$}}

\shortauthors{ASCHWANDEN 2015}
\shorttitle{Magnetic Energies from IRIS and AIA Data}

\begin{document}

\title{         Magnetic Energy Dissipation during the 2014 March 29 
		Solar Flare}

\author{        Markus J. Aschwanden$^1$	}

\affil{		$^1)$ Lockheed Martin, 
		Solar and Astrophysics Laboratory, 
                Org. A021S, Bldg.~252, 3251 Hanover St.,
                Palo Alto, CA 94304, USA;
                e-mail: aschwanden@lmsal.com }

\begin{abstract}
We calculated the time evolution of the free magnetic energy during the 
2014-Mar-29 flare (SOL2014-03-29T17:48), the first X-class flare detected 
by IRIS.  The free energy was calculated from the difference between the 
nonpotential field, constrained by the geometry of observed loop structures,
and the potential field. We use AIA/SDO and IRIS images to delineate the 
geometry of coronal loops in EUV wavelengths, as well as to trace magnetic 
field directions in UV wavelengths in the chromosphere and transition region. 
We find an identical evolution of the free energy for 
both the coronal and chromospheric tracers, as well as agreement between 
AIA and IRIS results, with a peak free energy of $E_{free}(t_{peak})
\approx (45 \pm 2) \times 10^{30}$ erg, which decreases by an amount of 
$\Delta E_{free} \approx (29 \pm 3) \times 10^{30}$ erg during the flare decay 
phase. The consistency of free energies measured from different EUV and UV 
wavelengths for the first time here, demonstrates that vertical electric 
currents (manifested in form of helically twisted loops) can be detected 
and measured from both chromospheric and coronal tracers.
\end{abstract}


\section{		INTRODUCTION				}

Since magnetic reconnection processes are believed to be the primary
source of energy for producing solar flares, filament eruptions,
and coronal mass ejections (e.g., Priest 1982; 2014), 
the measurement of the dissipated
magnetic energy provides a key parameter in the understanding of 
the underlying physics. The dissipated magnetic energy is thought to
represent an absolute upper limit to all secondary energy conversions, 
such as thermal, nonthermal, radiative, and kinetic energies. The
measurement of the dissipated magnetic energy requires a reliable method
of calculating the evolution of the nonpotential magnetic field 
during a flare. The difference between the 
nonpotential $E_{np}(t)$ and the potential energy $E_p(t)$ 
is the maximum free energy, i.e., $E_{free}(t)=E_{np}(t)-E_{p}(t)$, which
provides an upper limit on the total dissipated energy in a flare.

There are two fundamentally different methods to calculate the
nonpotential energy: (i) using a nonlinear force-free field (NLFFF)
code that extrapolates from the 3D vector magnetic field at the
photospheric boundary (which we call the PHOT-NLFFF method;
e.g., Wiegelmann 2004), and 
(ii) by forward-fitting of a NLFFF approximation
to the observed geometry of coronal loops, using a line-of-sight (LOS)
magnetogram to constrain the potential field (which we call the
COR-NLFFF method; Aschwanden 2013a). 
For the first method exist about a dozen of
NLFFF codes, which have been compared and showed a large scatter
of the free energy (Schrijver et al.~2006, 2008). Moreover, the most 
severe problem of PHOT-NLFFF codes is their underlying assumption that
the photospheric boundary is force-free (DeRosa et al.~2009),
although attempts have been made to bootstrap the force-freeness
of the photospheric boundary by a ``pre-processing method'' 
(Wiegelmann et al.~2006). The question arised: 
{\sl Can we improve the pre-processing of photospheric vector
magnetograms by the inclusion of chromospheric observations?}
(Wiegelmann et al.~2008). In contrast, the COR-NLFFF
code circumvents the non-force-free photosphere by fitting a
quasi-force-free solution to loops in force-free regions of the corona.  
Using this second (COR-NLFFF) method, the dissipated magnetic energies 
could be determined for 172 major (GOES M and X-class) flare events,
yielding dissipated energies that amount to a fraction $E_{diss}/E_p
\approx 1\%-25\%$ of the potential energy, where the potential
field covers a range of $E_p \approx 10^{31}-10^{33}$ erg for large
(M and X-class) flares (Aschwanden, Xu, and Jing 2014; 
Emslie et al.~2012).   

The accuracy in the calculation of free energy crucially depends
on the force-freeness of the boundary field or fitted loops.
While the solar corona is believed to be force-free in most places,
major parts of the transition region, 
the chromosphere, and the photosphere are dominated by regions with
a high plasma-$\beta$ (i.e., the ratio of the thermal to magnetic energy)
that is larger than unity (e.g., Gary et al.~2001), which can enable
cross-field electric currents that disturb the force-freeness 
condition.  Measurements of
the chromospheric vector field and application of the virial theorem
demonstrated that the photosphere and lower chromosphere is not 
force-free, while it becomes force-free at an altitude of
$h \gapprox 400$ km (Metcalf et al.~1995). Attempts to improve the 
accuracy of NLFFF solutions have been made by using H$\alpha$ observations 
(Wiegelmann et al.~2008), which outline loop-like or ribbon-like structures 
in the chromosphere. Here we apply the COR-NLFFF
method to images obtained in coronal EUV wavelengths (with AIA),
as well as (for the first time) to images obtained in the chromosphere and
transition region in UV wavelengths (with IRIS and AIA), and we compare the 
evolution of the free energy inferred in both height regimes. 

\section{ 	OBSERVATIONS AND MEASUREMENTS			}

\subsection{	AIA, HMI, and IRIS Observations			}

We perform modeling of the nonpotential magnetic field for the
flare on 2014 March 29, 17:35-17:54 UT, classified as a GOES 
X1.0-class event, which occurred in the NOAA active region 12017, 
located at heliographic position N11W32. It was the first X-class
flare observed by IRIS and has been declared as the 
{\sl ``best-ever observed flare''} (NASA press release of 2014 May 7).
Recent studies on this flare deal with the origin of a sunquake
(Judge et al.~2014), the hydrogen Balmer continuum emission during
the flare (Heinzel and Kleint 2014), and spectroscopy at subarcsecond 
resolution (Young et al.~2015).

We are using images from the {\sl Atmospheric
Imager Assembly (AIA)} (Lemen et al.~2012) onboard the {\sl Solar
Dynamics Observatory (SDO)} (Pesnell et al.~2011), and slit-jaw
images (SJI) from the {\sl Interface Region Imaging Spectrograph
(IRIS)} (De Pontieu et al.~2014). A list of the analyzed
wavelengths is given in Table 1, which we group into three
wavelength sets used in three independent data analysis runs: 
(i) IRIS-UV with wavelengths 1400 and 2796 \ang\ 
that probe the chromosphere and transition region, 
(ii) AIA-UV wavelengths 304 and
1600 \ang\ that probe the transition region also, and (iii) AIA-EUV
wavelengths 94, 131, 171, 193, 211, 335 \ang , that probe the corona.
In addition we use magnetograms from the {\sl Helioseismic and 
Magnetic Imager (HMI)} (Scherrer et al.~2012) onboard SDO. For our 
analysis we use a cadence of 3 minute in all wavelengths, which
yields 26 time steps during an interval that covers the entire
flare duration plus 0.5 hrs margins before and after (i.e.,
17:05-18:24 UT). The AIA images have a pixel size of $0.6\arcsec$,
the HMI magnetograms have a pixel size of $0.5\arcsec$, and
IRIS SJI images have a pixel size of $0.166\arcsec$.
For the field-of-view of the analyzed subimages we use a
square with a length of 0.2 solar radii, centered at location
N11W32, adjusted to solar rotation tracking during the observed 
time interval.

\subsection{		Data Analysis Method			}

We determine the evolution of the free energy $E_{free}(t)$ during
the flare episode by applying the COR-NLFFF code, for which
the theoretical framework is given in Aschwanden (2013a), numerical
tests in Aschwanden and Malanushenko (2013), and the determination
of the free energy in Aschwanden (2013b). The theoretical concept
of the COR-NLFFF code is based on an analytical NLFFF solution of
buried magnetic charges, each one having a variable vertical current
and an associated helical twist of the field lines. The analytical
solution is divergence-free and force-free with an accuracy of 
second order of the azimuthal (non-potential) magnetic field component 
${\bf B}_{\varphi}({\bf x})$.
The performance of the COR-NLFFF code includes three tasks:
(1) A decomposition of the LOS component of the HMI magnetogram into
a finite number of magnetic charges that yield the potential
field solution ${\bf B}_p({\bf x})$ of an active region, 
(ii) automated loop tracing in AIA and IRIS images with the
OCCULT-2 code (Aschwanden et al.~2008, 2013b; Aschwanden 2010), 
separately executed for wavelength sets with
coronal and transition-region temperatures,
and (iii) forward-fitting of the non-potential (force-free) 
$\alpha$-parameters to the coronal magnetic field by optimizing
the 2D-misalignment angles between the theoretical model of the 
nonpotential field ${\bf B}_{np}({\bf x})$ 
and the observed geometry of coronal loops. The NLFFF fitting procedure
follows closely the code version applied in the most recent statistical 
study of 172 major flares (Aschwanden et al.~2014), while the
application to loop structures observed in the transition region,
imaged by IRIS and AIA, represents a new experimental step explored
for the first time here. 

The standard control parameters of the COR-NLFFF code are given 
in Table 2 of Aschwanden et al.~(2014). In the present experiment
we used slightly different settings to optimize the results 
obtained from the IRIS images, in the sense of maximizing the
number of detected structures and minimizing the number of non-loop 
features: 
curvature radius $r_{min}=15$ (or 10) pixels for EUV (or UV) images,
loop length limit $l_{min}=r_{min}$,
no gaps in the loop structures $n_{gap}=0$, 
flux profile rippledness $q_{ripple} \le 0.4, 0.6, 0.8$,
flux threshold $q_{thr} \ge 3.0$ standard deviations,
number of magnetic sources $n_{mag}=100$,
proximity for the separation of loop footpoints $d_{prox} \le 3$ FWHM,
number of iterations $10 \le n_{iter} \ge 20$,
limit of loop structures per wavelength $n_{loop} \le 200$,
number of loop segments $n_{seg}=7$,
and maximum altitude of $h=0.05$ (0.20) solar radii for IRIS (AIA).
Differences between AIA and IRIS images are mostly the spatial
resolution, the flux contrast, and the morphology of wavelength-specific 
features.

\section{		RESULTS 				}

A snapshot of the results of the magnetic modeling of the 2014 Mar 29
flare is shown in Figs.~1 and 2, at 17:53 UT, at the
time of the flare peak. Original images
of the flare region are shown on a logarithmic flux scale in Fig.~1
(left), juxtaposed to the highpass-filtered images (Fig.~1, right)
that have been used for automated loop detection (red curves in
Fig.~1 right panels).  Fig.~2 shows an AIA EUV image sensitive to
coronal temperatures (211 \ang\ ), an AIA 304 \ang\ (Fig.~2, top left) 
and an IRIS SJI image at 2796 \ang\ (Fig.~2 middle left), both being
sensitive to chromospheric and transition region temperatures. The
loop structures traced in these images, found with the automated 
pattern recognition code OCCULT-2, are indicated in red. The structures
detected in IRIS images are generally shorter segments than those
detected in AIA images, which is partly due to a different morphology, 
and partly due to the 4 times higher spatial resolution of IRIS.

The geometric 2D coordinates of the automated
loop tracings constrain the best fit of the NLFFF solution, 
which is shown together with the HMI magnetogram
in Fig.~2 (middle right panel), (from which the NLFFF code uses
only the LOS component). We see that the active region contains
closed loops (blue curves) in the eastern dipolar part, while there
are mostly open field lines in the western dipolar spot group,
except for a small bipolar arcade in the core, where most of the
flare action occurs. Integrating the free magnetic energy in every
LOS, we created a free energy map (Fig.~2 bottom right),
which reveals that the largest amount of free energy is concentrated
in a semi-circular configuration surrounding the penumbra of the main
western sunspot. We show the 50\% contours of the free 
energy map $E_{free}(x,y,t)$ at 17:53 UT, shortly after the flare peak, 
overlaid on the magnetogram (Fig.~2 middle right panel, red contour), 
which is essentially identical with
the location of the dissipated magnetic energy (i.e., the difference
of the free energy between flare start and end). The eastern part
of the semi-circular energy dissipation structure is cospatial with
the location where 30-70 keV non-thermal hard X-ray emission was
detected with RHESSI, as well as enhanced (hydrogen Balmer) white 
light continuum with IRIS 1400 \ang\ (see Fig.~1 of Heinzel and
Kleint 2014). In any case, the spatial map of the free energy
localizes the footpoint areas of the flare loops that undergo
magnetic reconnection. We find essentially identical free energy
maps for coronal (AIA EUV wavelengths) and chromospheric or
transition region (AIA UV and IRIS wavelengths) loop tracers,
which implies that the magnetic information of the nonpotential
field $B_{np}(x,y,t)$ and free energy $E_{free}(x,y,t)$ 
is imprinted in field-aligned structures that can be observed 
in the chromosphere, the transition region, and in the corona.

A new result of this study is that we extended nonpotential
magnetic modeling from the corona down to the transition region
and chromosphere, by calculating NLFFF solutions from three
different wavelength sets, i.e., AIA-EUV, AIA-UV, and IRIS-UV.
We show the resulting evolution of the magnetic energies calculated
from these three wavelength sets in Fig.~3, performed in time steps
of $\Delta t=0.05$ hrs (3 min) during the flare time interval 
(with 0.5 hrs margin before and after). Error bars of the free
energy measurements are estimated from 3 different loop selection
criteria (defined by the level of flux variability along each
detected loop segment, $q_{ripple}=0.4, 0.6, 0.8$).
We find that the free energy peaks at a value of
$E_{free} \approx (45 \pm 2) \times 10^{30}$ erg 
for all three wavelength sets. All three evolutions of free 
energies peak consistently within a few minutes after the 
GOES-based flare peak time (Fig.~3, bottom panel). Both the
AIA-EUV and AIA-UV wavelength sets exhibit a step-wise drop 
of the free energy within 10-20 min after the flare peak, while
this step could be verified in the IRIS data partially only, due 
to an observing sequence that unfortunately ends at 17:54:19 UT.
Nevertheless, both the coronal and the chromospheric
data exhibit a consistent drop in free energy, by an average
amount of $\Delta E_{free} \approx (29 \pm 3) \times 10^{30}$ erg, 
which indicates that the helically twisted magnetic field lines
relax to a state of lower twist in both the corona and the
chromosphere, because the COR-NLFFF solution is most sensitive
to vertical currents and the associated helical twist around
vertical twist axes. This is the first quantitative measurement 
that demonstrates in both the chromosphere and corona that the 
dissipated magnetic energy in a flare is caused by untwisting of 
(helically-twisted) stressed field lines.

\section{ 		DISCUSSION 			}				 
\subsection{	Chromospheric Magnetic Field Tracers    }

Coronal loops are most conspicuously seen in EUV wavelength
images with a temperature sensitivity in the $T_e \approx 1-2$ MK
temperature range, such as in the Fe IX (171 \ang ) and in the
Fe XII (193 \ang ) line, and to a lesser extent in the weaker 
Fe XIV (211 \ang ) and Fe XVI (355 \ang ) lines (Table 1),
which all are most useful in constraining theoretical magnetic field
models, such as calculated with the COR-NLFFF code (Aschwanden 2013a) 
or with a Grad-Rubin NLFFF method (Malanushenko et al.~2014).
In the temperature regime of the photosphere, chromosphere, and
transition region, which ranges from $T_e=5000$ K to $T_e \lapprox
1$ MK, we expect to see cooler loops or the footpoints of hotter
coronal loops, which manifest themselves as a reticulated ``moss'' 
structure (Berger et al.~1999; De Pontieu et al.~1999). 
These chromospheric structures appear 
to be more irregular and fragmented than the smooth curvi-linear loop 
structures seen in the corona. Experiments to use the directions
of H$\alpha$ fibrils to constrain NLFFF solutions have been undertaken
by Wiegelmann et al.~(2008), a method that became to be known as
{\sl pre-processing}, which supposedly makes the photospheric 3D 
vector field more force-free and provides then a more suitable 
near force-free boundary for NLFFF extrapolation methods.
Due to regions with high plasma $\beta$-parameters and 
non-magnetic forces in the chromosphere, one cannot expect to model 
the chromospheric field correctly under the force-free assumption, 
neither with PHOT-NLFFF nor with COR-NLFFF codes.

Alternatively, we explore here for the first time the method of
using (automatically detected) chromospheric structures to measure
the local directivity of the magnetic field, which is then used
to constrain a nonpotential magnetic field (COR-NLFFF) solution in
an active region during a major flare. We use UV images in several 
wavelengths, such as the He II (304 \ang ) and C IV (1600 \ang ) lines
from AIA/SDO, and the Mg II (2796 \ang ) and
Si IV (1400 \ang ) lines from IRIS, which all exhibit some segments of
loop-like features in the chromosphere and transition 
region, and thus may contain some information on the local magnetic field
direction. Since the OCCULT-2 code is designed to detect
curvi-linear loop structures with large curvature radii, the
more irregular and inhomogeneous structures detected in the 
transition region and chromosphere are more challenging for
automated detection of magnetic field directions, and thus
it is not clear how useful they are for magnetic field modeling.
However, our experiment demostrates that we detect a consistent
evolution of the free energy $E_{free}(t)$ during the investigated
flare in both coronal EUV and chromospheric UV wavelengths
(Fig.~3), within the uncertainties of the measurements. The biggest 
challenge is to distinguish between contiguous field-aligned structures 
(segments of loops or filaments), curvi-linear aggregations
of ``moss''-like structures, and curved flare ribbons.

\subsection{	Energy Dissipation in Flares		}

The coronal and/or chromospheric feature tracing provides field
directions, which can be used to calculate the nonpotential magnetic
field ${\bf B}({\bf x})$ in a computation box that encompasses an
active region or flare region. The volume integral of the free
energy, $E_{free}(t) = (1/8\pi ) \int B_{np}(x,y,t)^2-B_p(x,y,t)^2
\ dx \ dy$, is ideally expected to have a higher level of free energy 
before the flare, and to drop to a lower level after the flare 
(e.g., Jing et al.~2009). We indeed find
a step-wise decrease of the free energy after the flare, which defines
the total dissipated magnetic energy during the flare, and is found
to have a value of $\Delta E_{free} = E_{free}(t_{start})-E_{free}(t_{end})
\approx (29 \pm 3) \times 10^{30}$ erg here. This value is indeed typical for an
X-class flare (Aschwanden et al.~2014). The apparent increase
of the free energy before the flare peak time has been interpreted as
an illumination effect, caused by chromospheric evaporation that fills up
flare loops, whereby their helical twist becomes visible
(Aschwanden et al.~2014). The dissipated magnetic energy is an 
important physical parameter for flare models, because it sets rigorous 
upper limits on other secondary flare energy conversions
(thermal and nonthermal energy, kinetic energy of CMEs, etc.), it
constrains the number problem for particle acceleration, and
constrains physical scaling laws for magnetic reconnection processes.
It is therefore imperative to establish a reliable method for the
determination of nonpotential magnetic energies. A comparison
of potential, nonpotential, and free energies between PHOT-NLFFF
and COR-NLFFF codes has been conducted in a recent statistical study,
where it was found that the dissipated flare energies determined with 
PHOT-NLFFF and COR-NLFFF methods disagree up to a factor of 10
(Aschwanden et al.~2014). At this point it is not clear what
the largest source of uncertainties is. It is possible that the
pre-processing of photospheric vector field data in the PHOT-NLFFF
method introduces an over-smoothing and leads to an underestimation 
of magnetic energies (Sun et al.~2012). Alternatively, mis-detections
or sparseness of field-aligned structures with the OCCULT-2 code could 
spoil the optimization of force-free $\alpha$-parameters in the 
COR-NLFFF code.
However, the consistent result of the evolution of the free energy
found in this study from both coronal and chromospheric images yields
an independent test and corroboration of the COR-NLFFF method.

\section{		CONCLUSIONS			}

We calculated the time evolution of the free energy $E_{free}(t)$
during the 2014-Mar-29 flare, the first X-class flare detected by IRIS.
We used HMI/SDO data to compute the potential field,
and AIA/SDO and IRIS images to delineate the geometry of coronal loop
segments in EUV wavelengths, as well as to delineate magnetic field 
directions from automatically traced structures seen in the transition 
region and chromosphere in UV wavelengths. 

The major result of this study is that we find a similar 
evolution of the free energy for both coronal and chromospheric
structures, peaking at a free energy of $E_{free}(t_{peak})
\approx (45 \pm 2) \times 10^{30}$ erg, which decreases by an amount of
$\Delta E_{free} \approx (29 \pm 3) \times 10^{30}$ erg in the flare 
decay phase, and thus represents the total magnetic energy dissipated
during this X-class flare. The consistency of magnetic energy measurements 
from both coronal and chromospheric tracers represents an independent
test and corroboration of the COR-NLFFF method. The COR-NLFFF code
provides also maps of dissipated flare energies, which enables us
to localize and map out magnetic reconnection regions during solar
flares in great detail, which is the subject of future studies.

\bigskip
\acknowledgements
We acknowledge helpful comments from the referee, the LMSAL Team, 
and guidance for using IRIS data by Bart De Pontieu.
Part of the work was supported by the
NASA contracts NNG04EA00C of the SDO/AIA instrument and
NNG09FA40C of the IRIS mission.


\clearpage


\begin{deluxetable}{llll}
\tabletypesize{\footnotesize}
\tablecaption{Instruments and wavelengths used in the present analysis.}
\tablewidth{0pt}
\tablehead{
\colhead{instrument}&
\colhead{Wavelength}&
\colhead{Atomic Lines}&
\colhead{Temperature range}\\
\colhead{}&
\colhead{$[\ang ]$}&
\colhead{}&
\colhead{$\log(T [K])$}}
\startdata
\underbar{Chromosphere and}	    & & &       \\ 
\underbar{Transition Region:}	    & & &	\\ 
IRIS	& 2796          & Mg II h/k & 3.7-4.2	\\ 
IRIS	& 1330		& C II      & 3.7-7.0 	\\
IRIS	& 1400          & Si IV     & 3.7-5.2	\\
AIA	&  304          & He II	    & 4.7	\\
AIA	& 1600          & C IV	    & 5.0	\\
\underbar{Corona:}	& & 	    &		\\
AIA     &  171          & Fe IX     & 5.8	\\
AIA 	&  193          & Fe XII, XXIV& 6.2, 7.3\\
AIA     &  211          & Fe XIV    & 6.3	\\
AIA 	&  355          & Fe XVI    & 6.4	\\
AIA     &   94          & Fe XVIII  & 6.8	\\
AIA	&  131          & Fe VIII, XXI& 5.6, 7.0\\
\enddata
\end{deluxetable}
\clearpage


\begin{figure}
\plotone{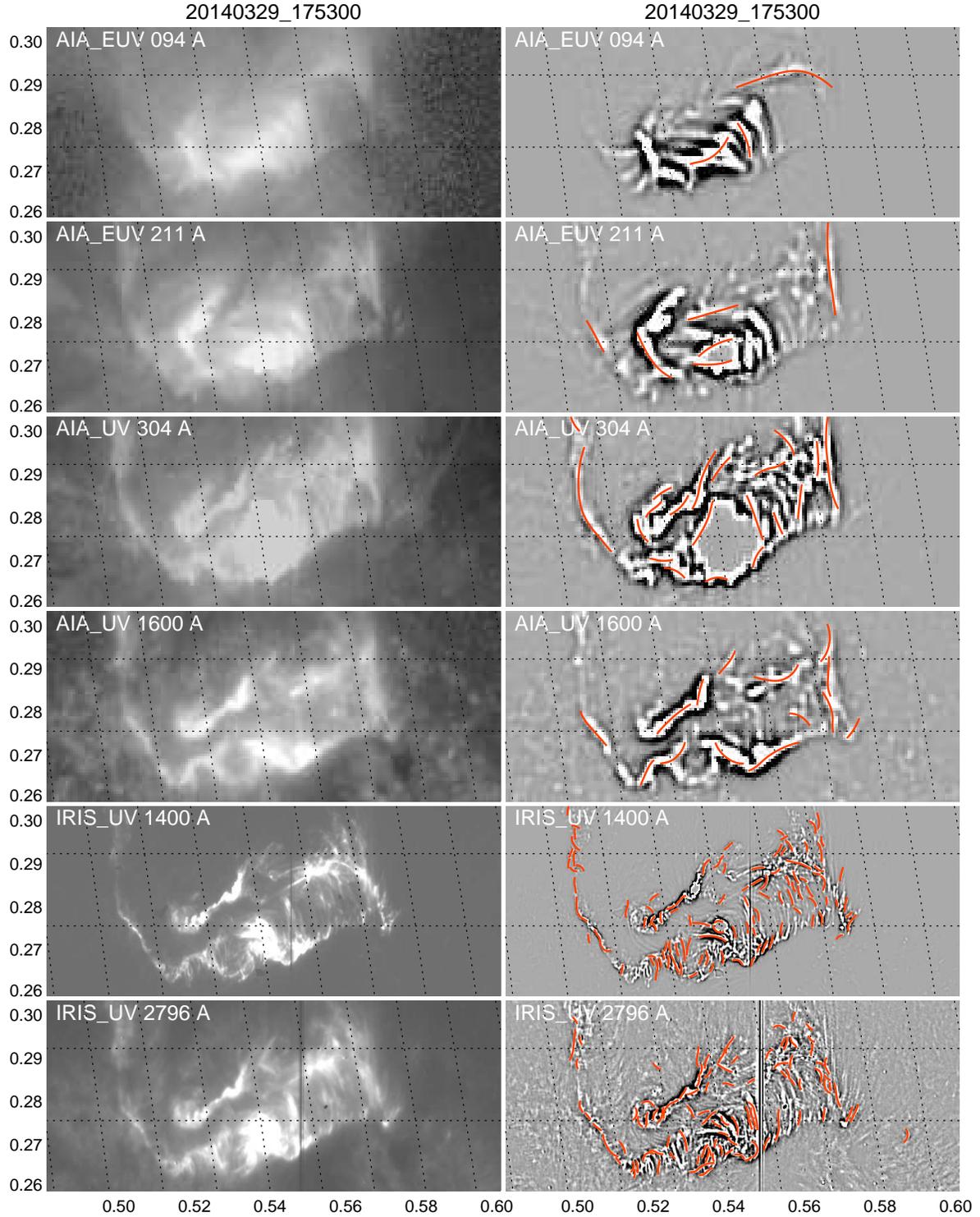}
\caption{Original EUV and UV images are shown in intensity
(left column) and as highpass-filtered fluxes (right column),
with the overlaid loop segments automatically traced by OCCULT-2
(red curves in right panels). The six wavelengths 
include the least-saturated EUV images (94 and 211 \ang ) 
of AIA, UV images of AIA (304 and 1600 \ang ), and UV slit-jaw 
images of IRIS (1400 and 2796 \ang ). Note the difference in 
spatial resolution ($0.6\arcsec$ for AIA and $0.166\arcsec$ 
for IRIS).}
\end{figure}
\clearpage

\begin{figure}
\plotone{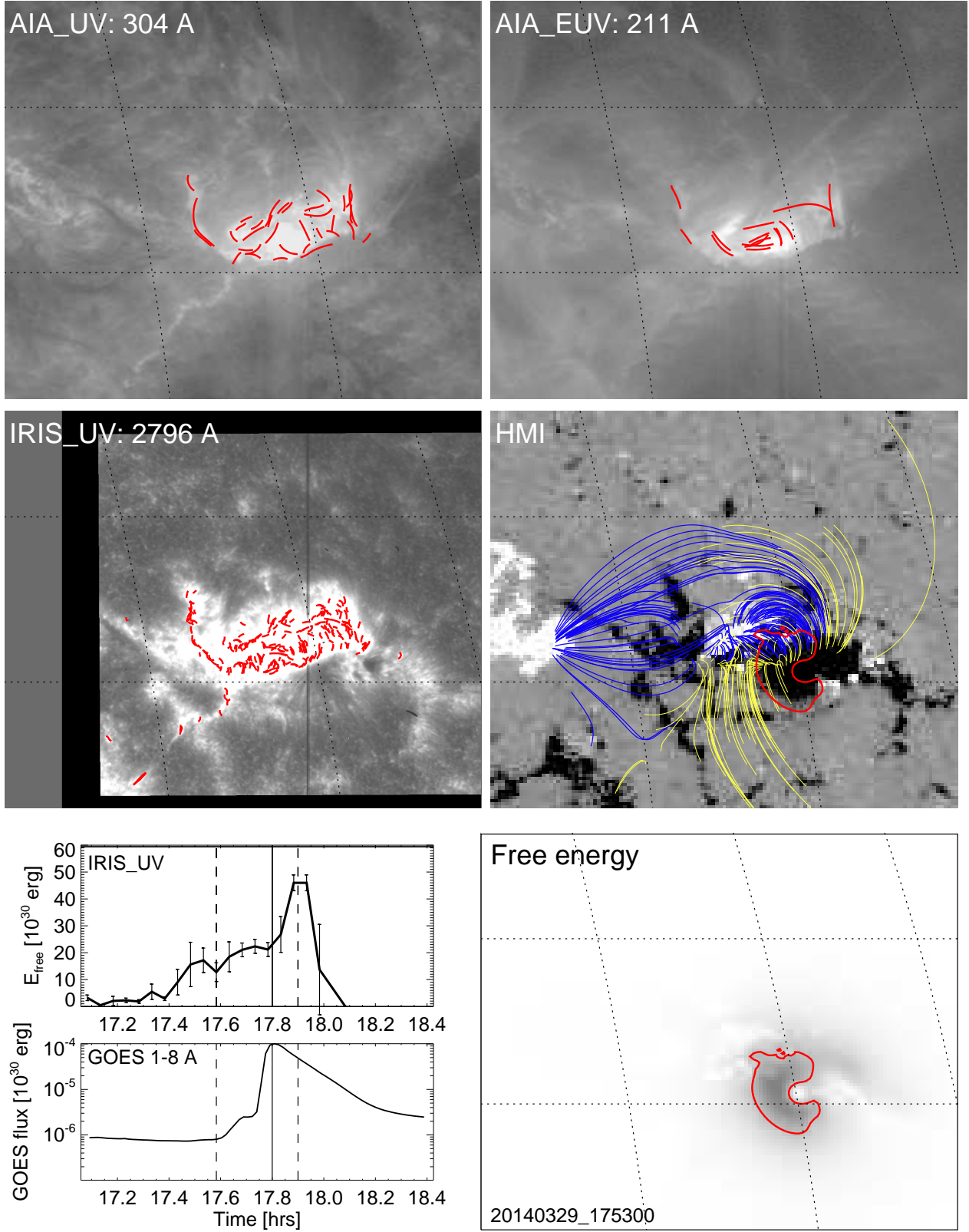}
\caption{Highpass-filtered images from AIA 304 \ang\ (top left),
AIA 211 \ang\ (top right), and IRIS 2796 \ang\ (middle left), with
overlaid loop tracings using an automated pattern detection code
(OCCULT-2: red curves), HMI magnetogram (middle right; with range
$-1472 \le B_z \le 972$ G), with overlaid 
forward-fitted nonlinear force-free field lines (COR-NLFFF), 
delineating closed loops (blue curves) and open field lines 
(yellow curves), the spatial distribution of the free energy 
(bottom right), with a 50\% contour level, the temporal evolution 
of the free energy measured from IRIS-UV (third left panel), 
and the GOES 1-8 \ang\ light curve (bottom left panel).} 
\end{figure}
\clearpage

\begin{figure}
\plotone{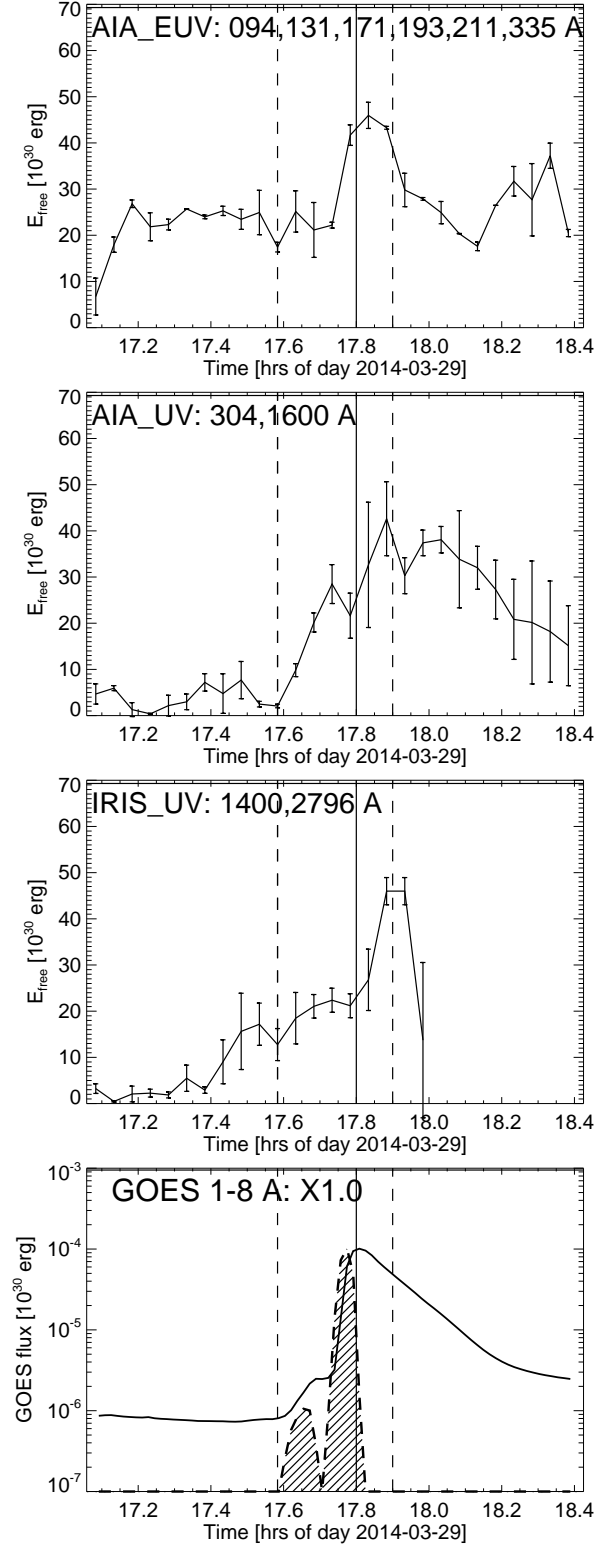}
\caption{Time evolution of the free energy as measured from
AIA in EUV wavelengths (top panel), from AIA UV wavelengths 
(second panel), from IRIS UV wavelengths (third panel),
along with the GOES 1-8 \ang\ flux (solid linestyle in bottom panel)
and GOES time derivative (dashed linestyle in bottom panel).
The errors of the free energies are estimated from 3 different
loop selections (fluctuation thresholds of 40\%, 60\%, and 80\%.) 
The start, peak, and end time of the GOES event is indicated
with vertical lines.}
\end{figure}
\clearpage

\end{document}